\documentclass{article}
\usepackage{amsmath,graphicx,mlspconf}
\usepackage{float}
\usepackage[belowskip=-15pt,aboveskip=10pt]{caption}

\copyrightnotice{978-1-7281-6662-9/20/\$31.00 {\copyright}2020 Crown}

\toappear{2020 IEEE International Workshop on Machine Learning for Signal Processing, Sept.\ 21--24, 2020, Espoo, Finland}

\title{Frequency Domain-based Perceptual Loss for Super Resolution}
\name{Shane D. Sims}
\address{University of British Columbia \\
        Department of Computer Science \\
        Vancouver, BC, Canada}

\begin{document}

\maketitle
\begin{figure*}[t]
\centering
  \includegraphics[width=2.0\columnwidth]{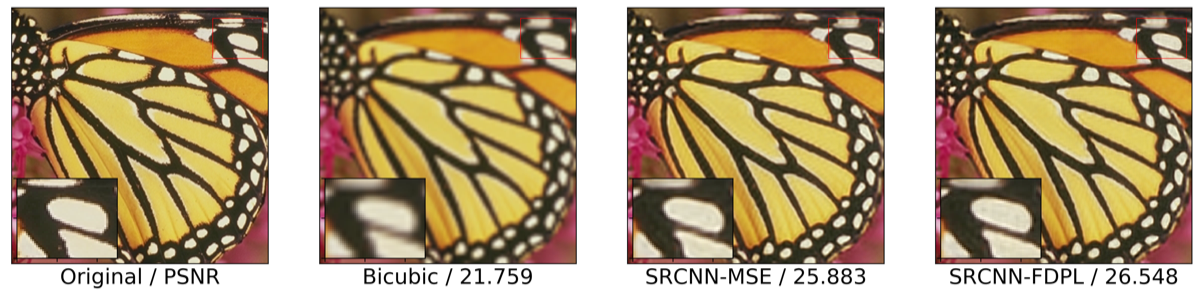}
  \caption{The image resolved by SRCNN trained with proposed FDPL (right) surpasses the image resolved by SRCNN trained with MSE pixel loss (second from right), in terms of PSNR (dB, higher is better) and perceptual quality. Upscale factor: 3.}~\label{fig:figure1}
\end{figure*}
\begin{abstract}
We introduce Frequency Domain Perceptual Loss (FDPL), a loss function for single image super resolution (SR). Unlike previous loss functions used to train SR models, which are all calculated in the pixel (spatial) domain, FDPL is computed in the frequency domain. By working in the frequency domain we can encourage a given model to learn a mapping that prioritizes those frequencies most related to human perception. While the goal of FDPL is not to maximize the Peak Signal to Noise Ratio (PSNR), we found that there is a correlation between decreasing FDPL and increasing PSNR. Training a model with FDPL results in a higher average PSRN (30.94), compared to the same model trained with pixel loss (30.59), as measured on the \textit{Set5} image dataset. We also show that our method achieves higher qualitative results, which is the goal of a perceptual loss function. However, it is not clear that the improved perceptual quality is due to the slightly higher PSNR or the perceptual nature of FDPL.
\end{abstract}
\begin{keywords}
Super Resolution, Perceptual Loss, Frequency Domain, Convolutional Neural Networks, Image Processing
\end{keywords}
\section{Introduction}
\label{sec:intro}

Single-image super resolution (SR) is a classic problem in computer vision, which seeks to recover a high-resolution output image from a given low-resolution input image \cite{park2003super}.
As low-resolution images are typically generated via processes that include down-sampling rows and columns of pixels, there may be no single solution to the problem. That is, many different high-resolution images can be down-sampled into the same low-resolution counterpart. The task of a SR model is to learn what should exist between any two rows or columns of a given low-resolution image; a difficult task considering the variation possible in even a small set of images. 


Since being approached using deep learning methods, improvements in super resolution performance have been driven mainly by innovations in model architecture. Few previous works focus on adjusting the loss function to correspond with what is most important for a given use case. Janocha and Czarnecki focused on this issue directly (for classification) and showed that the choice of loss function has a direct effect on learning dynamics and what is ultimately learned by the model \cite{janocha2017loss}. Most modern work on super resolution has addressed the difficulty of the task by training image transformation Convolutional Neural Networks (CNNs) of varying complexity using a per-pixel loss function. Such loss functions minimize the distance between a low-resolution image and its ground truth target, which is measured by either $\ell_1$ or $\ell_2$ norm-based functions. A notable exception to the use of pixel-loss is the work of Johnson \textit{et al.} \cite{johnson2016perceptual}, which uses a pre-trained loss network and compares the value at an intermediate activation layer of the loss network with the output of their image transformation network using mean squared error (MSE) between the two tensors. 

In this work, we too depart from using a pixel-loss and like Johnson \textit{et al.}, we construct a perceptual loss function to resolve images in a way that is more in line with human perception. Where we differ from the previous super resolution perceptual loss work (and all others before ours) is that we compute loss in the frequency domain, following a Fourier transform. Specifically, we use the Discrete Cosine Transform (DCT) \cite{ahmed1974discrete}, which enables us to use a perceptual model inspired by the JPEG image compression codec \cite{wallace1992jpeg}, the principles of which allow for excellent compression even when minimizing perceptible loss of quality. We use the same principles to guide our model to focus on resolving those parts of the image most important for human perception, by biasing the loss towards the frequency components lost during image downsampling and weighting these according to perceptual importance.
While our work is preliminary in nature, our experiments show that a model trained with FDPL achieves higher quantitative performance (measured by PSNR) than the same model trained with MSE pixel loss (see Fig. 1). We also believe that the FDPL trained model produces better qualitative results, though it is difficult to tell if this is due to the increased PSNR or because of the perceptual emphasis of the loss function. 

More specifically, the primary contribution of this work is the introduction and evaluation of Frequency Domain Perceptual Loss, a novel, model-agnostic perceptual loss function for image super resolution. 

\section{Related Works}
As a classic computer vision problem, super resolution has a wide body of related literature (ex. \cite{johnson2016perceptual, wang2018esrgan, kong2018image, dong2015image, glasner2009super, lim2017enhanced, yang2010image}), spanning back to at least the 1980s (ex. \cite{gross1986super, hummel1987deblurring, irani1990super}). Most related to our work are those that use CNNs for the task, which mostly use a pixel loss during training. Pixel loss is instantiated in two ways in the literature: as either the mean Euclidean distance or mean absolute value distance between the pixels of the output and target images (i.e. $\ell_2$ or $\ell_1$ norms, respectively). Regardless of the particulars of the loss function used, all are computed in the pixel domain. Our work is different from all previous works in this respect, as we calculate loss in the frequency domain, in order to use the appealing properties of having access to frequency coefficients.

\subsection{Pixel Loss for Super Resolution}
Most related to this work is the seminal work of Dong and Loy \cite{dong2015image}, which introduced Super Resolution CNN (SRCNN). This was the first published use of CNNs for super resolution and is still used as a benchmark in the most recent papers for the task, as well as in the comprehensive overview conducted by Yang \textit{et al.} \cite{yang2019deep}. SRCNN uses a simple three-layer architecture, each of which is grounded in a sub-task completed by previous non-CNN based approaches \cite{dong2015image}. SRCNN was trained by minimizing a per pixel-loss; the mean square error between the transformed image and the ground truth (or target) image. 

The current state of the art in super resolution are Fong and Fowlkes' Predictive Filter Flow network (PFF) \cite{kong2018image} and Wang \textit{et al.}'s Enhanced Super Resolution Generative Adversarial Network (ESRGAN) \cite{wang2018esrgan} (described below). PFF uses a CNN to predict near-optimal filter flow operations as the basis of their model. Another notable work is the Progressive Super Resolution (ProSR) network of \cite{wang2018fully}, which achieves near state of the art results while running 5 times faster than competing methods. ProSR focuses on high factor super resolution ($8\times$) and operates by progressively up-sampling the input image by $2\times$ at each level of the model. ProSR also adopts a GAN (as ProGanSR) to achieve impressive photo-realistic results at high super resolution factors. Training of ProSR is done by a variation of curriculum learning (a form of easy to hard progressive training) using an $\ell_1$ norm-based loss function. 

\subsection{Perceptual Loss for Super Resolution} Departing from strictly pixel loss-based training, Johnson \textit{et al.} use a perceptual loss function for super resolution \cite{johnson2016perceptual}. Their key intuition was that not all pixels are equally important for a human when judging the perceived quality of an image. As an extreme example, they point out that a perfectly resolved picture that is simply shifted one pixel in any direction will suffer from a very poor pixel loss. To achieve a perceptual loss, they use a pre-trained VGG-16 \textit{loss network} and compare the output value at an intermediate activation layer of the loss network when given the resolved image of the network being trained (on the one hand) with the output at the same layer when given the ground truth (on the other hand). Loss measured as the mean squared error (MSE) between the two resulting output tensors. 

In another seminal work on super resolution, Ledig \textit{et al.}  propose SRGAN, which combines the use of a GAN and a perceptual loss \cite{ledig2017photo}. Their loss is formulated by measuring ``content" reconstruction quality, which is also based on similarity to high-level features of a VGG network. Again, this loss is measured in the pixel domain. ESRGAN is an improvement of SRGAN, that achieves excellent qualitative and state of the art quantitative results by addressing some of the shortcomings specific to SRGAN \cite{wang2018esrgan}.

\section{Method}

We address the shortcomings of pixel-loss trained models by introducing the model agnostic Frequency Domain Perceptual Loss (FDPL) function. The key intuition of our method is that what makes a good super resolution model is the ability to do particularly well at restoring the high-frequency components of a given image that are within the range of human perception (see Fig. 1). Yet super resolution models trained on pixel loss target all frequencies equally: the low frequencies that are already given in the distorted image and the highest frequencies that are not perceptible to the human eye. If we can induce a model to focus on frequencies that are neither contained in the distorted image but are still important to human perception, then we will be able to achieve higher perceptual quality, even if this doesn't translate to a higher PSNR (which is directly related to reducing pixel-loss MSE). This is the goal of FDPL. 

There are two main components in the FDPL, which we explain in turn before presenting the loss function: the Discrete Cosine Transform (DCT) and the JPEG luminance quantization table and its use in image compression. 

The DCT is a Fourier transform that, like the Discrete Fourier Transform (DFT), transforms a signal into the frequency domain. Both the DCT and DFT perform a change of basis from the time basis (or pixel basis in our case) to the frequency basis. The main difference between the DCT and the DFT is the former's use of cosine wave bases, versus the sine wave bases used by the latter. Formally the 2D DCT (denoted from here on simply as $DCT$) is defined \cite{ahmed1974discrete} as:

$$\text{\small{DCT}}(\boldsymbol{X}_{jk}) \\=$$
$$\sum^{M-1}_{m=0}\sum^{N-1}_{n=0}x_{mn}\cos{(\pi\frac{j}{M}(n+\frac{1}{2}))}\cos{(\pi\frac{k}{N}(m+\frac{1}{2}))}$$

where $\boldsymbol{X}$ is an input image (i.e a matrix), $M$ and $N$ are the height and width of the image (respectively), $m$ and $n$ are the usual row and column indices so that $x_{mn}$ is the value at row $m$ column $n$, $0\leq j \leq M-1$, and $0\leq k \leq N-1$. 

The intuition behind the JPEG compression standard is an important component of our loss function. Briefly, JPEG operates by performing 2D DCT operations over all non-overlapping $8\times8$ patches in a given image, divides the resulting coefficients by a quantization table, rounds these results to induce sparsity, and then applies run length and Huffman encoding to obtain the final result \cite{wallace1992jpeg}. JPEG is an effective compression standard because it operates by removing the frequencies from an image that minimally harm perceived quality \cite{wallace1992jpeg}. Even with significant compression, perceived quality is minimally harmed. JPEG operates on the principle that higher frequencies are less perceptible to the human eye than are lower frequencies. Thus, to preserve image quality during compression, we should start with eliminating the highest frequency components (bottom right in left side Fig. 3) and continuing towards the lower frequency components (top left in Fig. 3) until the desired level of compression is reached. JPEG encodes this assumption in the values of its quantization tables. An example of a quantization table used on the Luminance channel (when the image is in YCbCr format) is shown in Figure 4. Here we see that the elements corresponding to higher frequencies have larger values, which will more quickly result in zeroed out DCT coefficients during compression following the rounding operation mentioned above. 
\begin{figure}[t]
\centering
  \includegraphics[width=0.75\columnwidth]{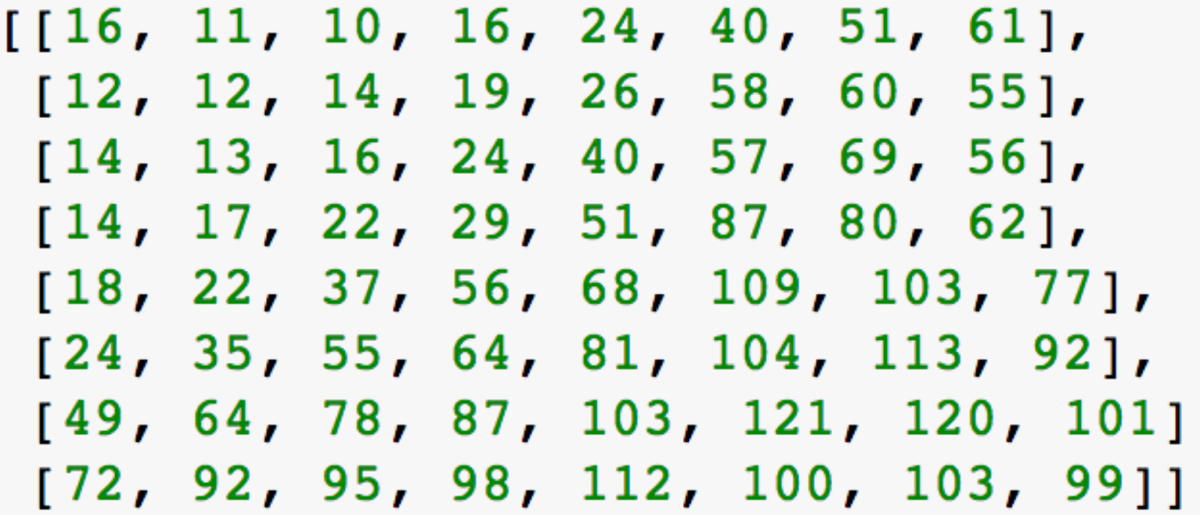}
  \caption{An example JPEG Luminance channel quantization table, encoding assumptions of the relative importance of frequency components for human perception.}~\label{fig:figure2}
\end{figure}

\begin{figure}[t]
\centering
  \includegraphics[width=0.45\columnwidth]{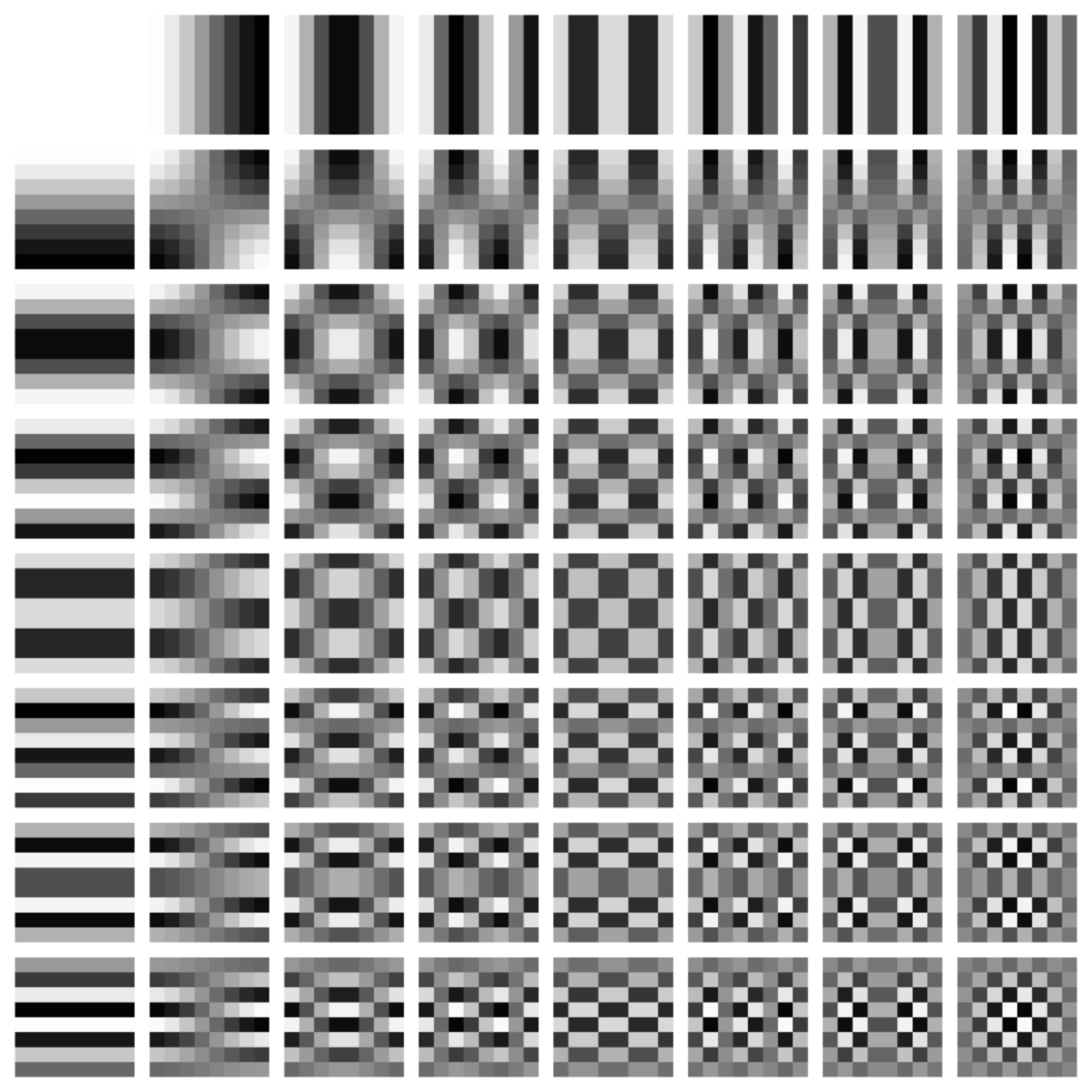}
  \includegraphics[width=0.51\columnwidth]{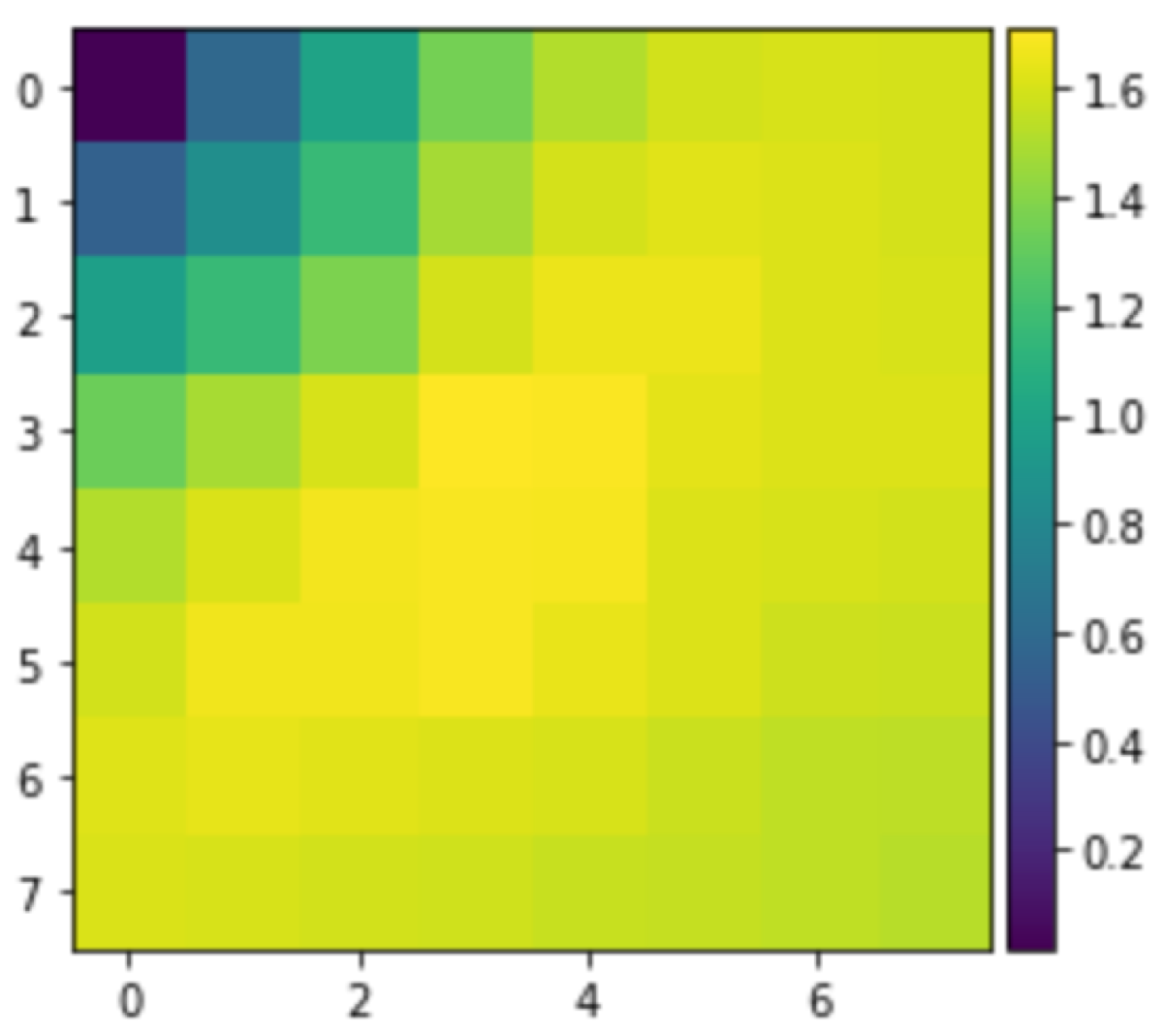}
  \caption{Left: a visualization of the DCT frequency components of an $8\times8$ image. Right: mean relative difference in frequency components between ground truth and distorted images in the training set.}~\label{fig:figure3}
\end{figure}

In this paper, we compute the 2D DCT function over $8\times8$ non-overlapping patches of the given image, unless stated otherwise. The output of this function is a matrix of the same dimensions as the input, where each element is a coefficient indicative of the strength of the corresponding basis element. In Figure 4 (left) we can see that the frequency of the components increases as we move from the top left to the bottom right. The resulting coefficient matrix can be interpreted as the amount of each frequency, that when added together, makes up the initial input image.

\subsection{Loss}
\begin{figure*}[t]
\centering
  \includegraphics[width=2.0\columnwidth]{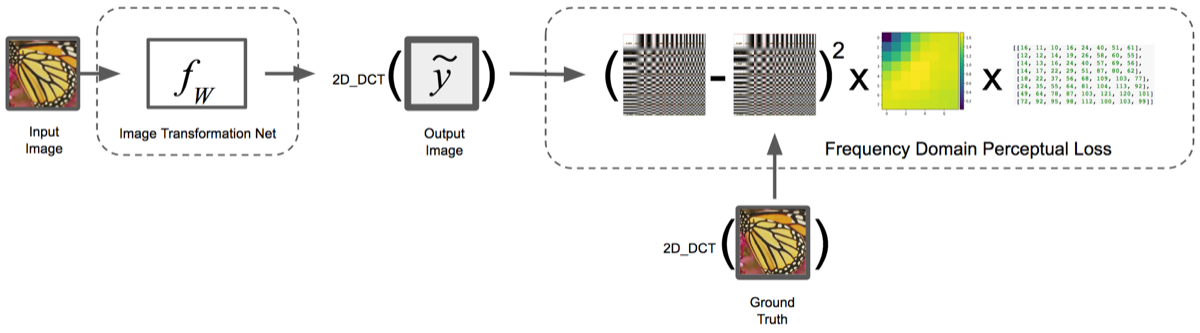}
  \caption{Visual overview of FDPL showing the squared difference in 8x8 DCT patches between target an output images being weighted by frequency importance, as determined by JPEG and relative difference frequencies computed over the training set.}~\label{fig:figure4}
\end{figure*}
We introduce Frequency Domain Perceptual Loss (FDPL) as a new loss function with which to train super resolution image transformation neural networks. Our method is model agnostic, so long as the model is trained via back-propagation. With the components described above (DCT and JPEG's quantization table) we can now define FDPL as follows:
\begin{align*}
    \text{FDPL}(\boldsymbol{y}, \Tilde{\boldsymbol{y}}) &= \lVert \text{DCT($\boldsymbol{y}$)} - \text{DCT($\boldsymbol{\Tilde{y}}$)}\rVert_2^2 \odot (\boldsymbol{1}\oslash\boldsymbol{q}) \odot \boldsymbol{d}
\end{align*}
where $\boldsymbol{y}$ and $\boldsymbol{\Tilde{y}}$ are the ground truth and output images (respectively), $\odot$ and $\oslash$ are the Hammarand (i.e. element-wise) multiplication and division operators (respectively), $\boldsymbol{q}$ is the quantization table (Fig. 3), and $\boldsymbol{d}$ is the mean relative difference matrix computed over the training set. Note that the difference matrix shown in Figure 4 supports our assertions that the low frequencies are given in the distorted image and that it is the medium-high frequencies (those across the main diagonal) that are lost in image down-sampling. Our loss thus induces the network to focus on these missing frequency components, according to their relative importance to human perception as encoded by $\boldsymbol{q}$. See Figure 2 for a visual overview of a system using FDPL.

\section{Experiments}
We investigate the impact of FDPL on super resolution performance by conducting two initial experiments, where the only independent variable is the loss function and all other variables are controlled. For these experiments, quantitative evaluation is done by measuring PSNR and mean structural similarity index (SSIM); two traditional metrics for measuring super resolution performance. Our main experiment compares the performance of the SRCNN model when trained with FDPL verses the MSE pixel loss described in \cite{dong2015image}. Our second experiment tests a variant of FDPL where the weights in $\boldsymbol{q}$ are transposed around the anti-diagonal (FDPL-AT from here on). The rationale for this latter experiment is that if restoring higher frequency elements is important for SR, then we should test our assumption that JPEG represents a good model of human perception by flipping importance to the highest frequencies. 

For our experiments, we follow the basic protocol in \cite{dong2015image}. We use the BSDS500 training (200 images) and test set (200 images) for our training set and the BSDS500 validation set (100 images) as our validation set. The training and target image pairs are prepared by taking $32\times32$ pixel patches with a stride of 13 over all of the images in the training and validation sets. This results in a training set of approximately 204800 image patches. To distort the input training images we downsample with bicubic interpolation by the super resolution factor. For all of our experiments, we use a super resolution factor of 3, which means that the output image will have $3\times$ the resolution of the input image. Input images are then blurred with a Gaussian kernel with a standard deviation of 1. Images are then up-sampled via bicubic interpolation by the super resolution factor as the sole pre-processing step. For simplicity, all experiments are done only on the Luminance channel of the YCbCr formatted images, as in \cite{dong2015image,johnson2016perceptual}.

Training of the SRCNN models is done via standard Stochastic Gradient Descent (SGD) with a batch size of 128, a learning rate of $1\times10^{-4}$ for the first two convolutional layers, and $1\times10^{-5}$ for the last. Models are implemented using PyTorch \cite{paszke2019pytorch} and trained on a single Nvidia GTX 1080 GPU for $15\times10^6$ back-propagation operations\footnote{In \cite{dong2015image} SRCNN is trained for $8\times10^8$ back-propagation operations. We found that we achieved good results for experimentation purposes much earlier.}. Each model takes approximately 3 days to train from scratch. Our code and experiments are made available on GitHub at \textit{https://github.com/sdv4/FDPL} to facilitate further research in this direction. 


%

\subsection{Results}
\begin{figure*}[ht]
\centering
  \includegraphics[width=1\columnwidth]{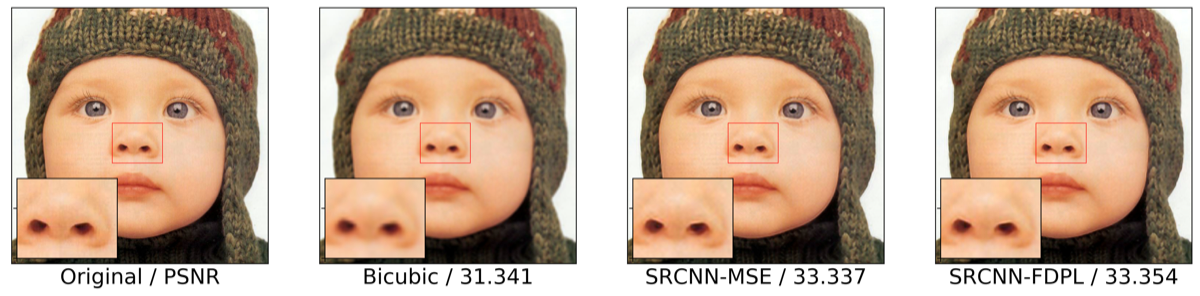}\hfill
  \includegraphics[width=1\columnwidth]{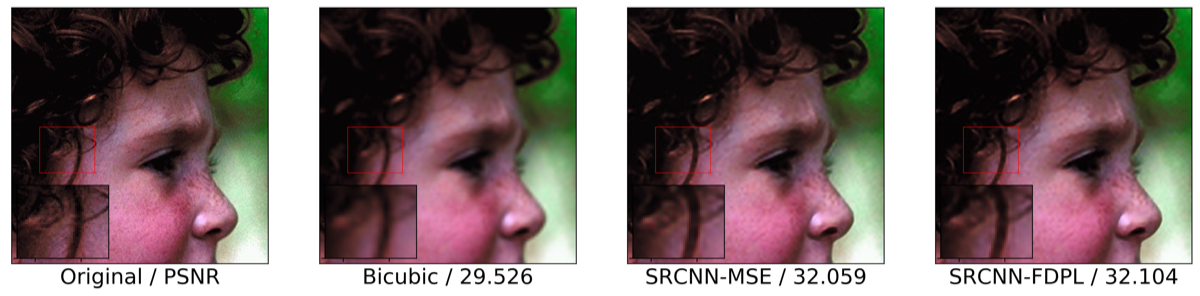}
   \\[\smallskipamount]
   \includegraphics[width=1\columnwidth]{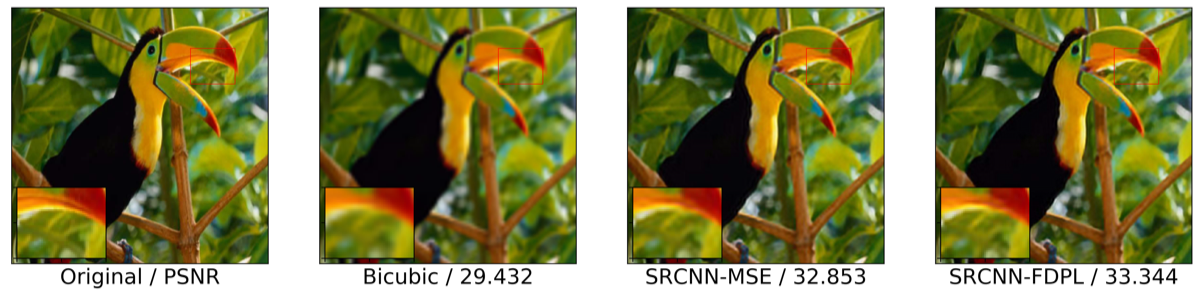}\hfill
  \includegraphics[width=1\columnwidth]{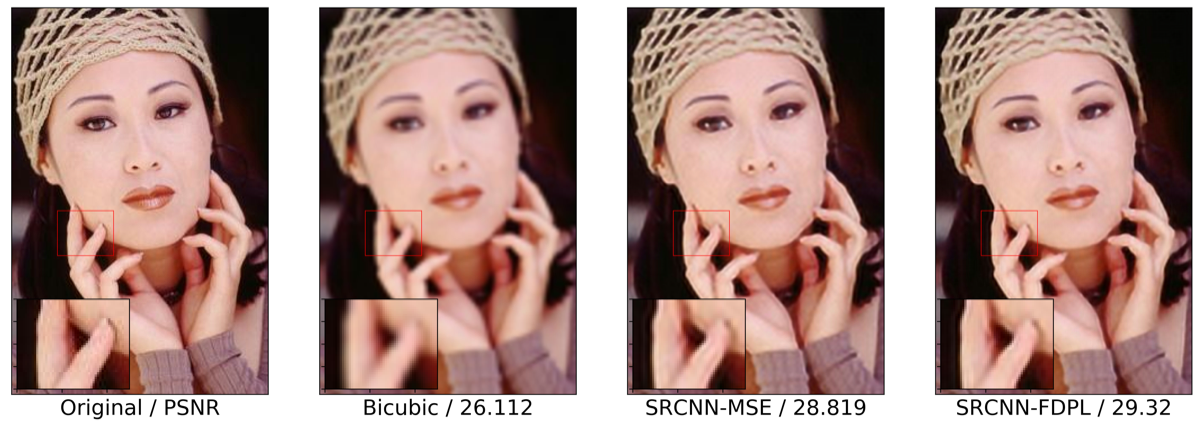}
  \setlength{\belowcaptionskip}{-1pt}
  \caption{The original, bicubic, SRCNN (MSE loss), and SRCNN (FDPL) versions of the \textit{baby, head, bird,} and \textit{woman} images from \textit{Set5}. Upscale factor: 3}~\label{fig:figure5}
\end{figure*}

\begin{figure}
\centering
  \includegraphics[width=1.0\columnwidth]{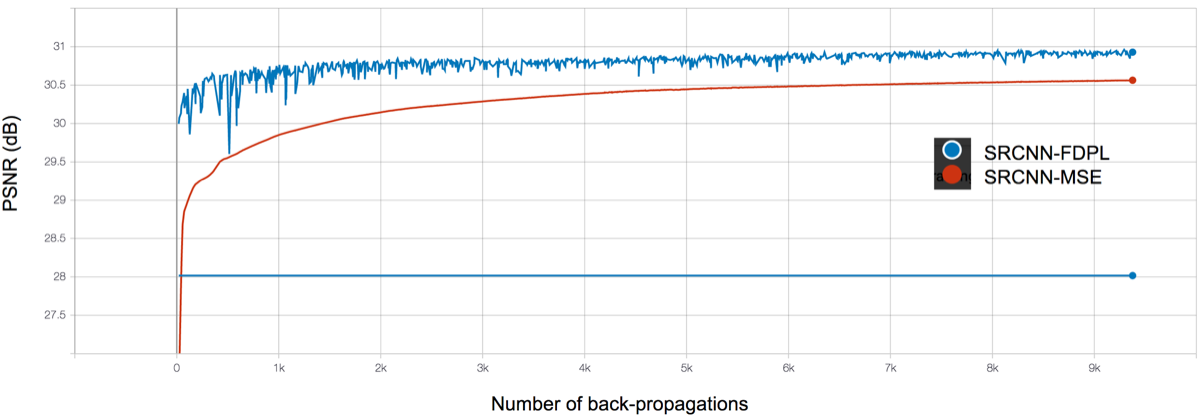}
  \caption{Average \textit{Set5} PSNR increases over training progression for SRNCC-FDPL (blue) and SRCNN-MSE (red). PSNR after distorting image shown as constant.}~\label{fig:figure6}
\end{figure}

While training SRCNN with MSE and FDPL, we monitor the average PSNR on the \textit{Set5} \cite{bevilacqua2012low} image set (as in \cite{dong2015image}). We can see in Figure 6 that, almost immediately, SRCNN trained with FDPL (SRCNN-FDPL) overtakes the PSNR achieved by SRCNN trained with MSE (SRCNN-MSE). While in both cases the average PSNR increases throughout the course of training, it is not smooth when trained with FDPL. This is an indication that while related to PSNR, FDPL is not optimizing this measure as directly as MSE does. At the conclusion of training, we see that using FDPL achieves a modest increase in average PSNR over MSE, giving empirical support to FDPL as a more suitable super resolution loss function. 

More important to us than quantitative performance is the perceived quality of the super resolved images. In Figure 5, we present a comparison of the images produced by the competing loss functions. While we believe that the images coming from SSRN-FDPL are more visually pleasing, this remains to be evaluated via a formal user study. At this point, we are also not sure if the higher perceived quality is a result of the perceptual qualities of FDPL, or because of the slightly higher PSNR it achieves; a question that requires further study. 


\begin{table}[t]
  \centering
  \begin{tabular}{l r r }
    {\textit{Loss}}
    & {\textit{PSNR}}
      & {\textit{SSIM}}\\
      \hline
    Bicubic (baseline) & 27.6340 & 0.8232\\
    MSE & 30.5917 & 0.8952 \\
    FDPL-AT & 30.5946  & 0.8988\\
    \textbf{FDPL} & \textbf{30.9399} & \textbf{0.9016}\\
  \end{tabular}
  \caption{Average PSNR and SSIM achieved by SRCNN \cite{dong2015image} when trained with pixel loss (MSE), FDPL with $q$ transposed across the anti-diagonal (FDPL-AT), and the proposed FDPL. }~\label{tab:table1}
\end{table}
\section{Conclusion}

In this paper, we introduced the Frequency Domain Perceptual Loss function for training image transformation networks for super resolution. We conducted preliminary experiments evaluating the efficacy of this loss function for single image SR, where we showed that its use in training SRCNN results in modest improvements in quantitative performance, as measured by PSNR and SSIM. We believe that the images resolved by the FDPL trained SRCNN are also of better perceptual (qualitative) quality. The main limitation of this result is that we can not say for certain that this qualitative improvement comes from the perceptually oriented FDPL function or simply as a result of the increased PSNR that it achieves. 

The areas of future work are numerous and in addition to expanding the results of this paper, are required to answer the open question just mentioned. More specifically, before training state of the art super resolution networks (including GAN-based approaches) with FDPL, we must first train SRCNN with more super resolution factors. To match the related super resolution literature, this will require training for $2\times, 4\times$, and $8\times$ upscale factors. We would also like to evaluate the mean relative difference matrix component in FDPL to provide further support for its use. Following the evaluation done in \cite{dong2015image}, we will also train our models with a subset of the ImageNet1K dataset in order to have a more diverse training set. 

Future work should also explore the performance of FDPL in the image compression artifact reduction task, where models and methods are similar to SR, but where our loss function may be more complementary, given the use of the DCT in many compression algorithms. A reasonable place to start would be with the work of Yu \textit{et al.} \cite{yu2016deep}, which is a slightly modified version of SRCNN for the artifact reduction task. We are currently undertaking this work and have obtained promising preliminary results. Additionally, we will investigate the use of the Discrete Fourier Transform as an alternative to the DCT, since it is known to distribute frequency information more equally across basis coefficients as compared to DCT, which concentrates information in the lower frequency elements. 

Finally, in this paper, we only used a 5 image test set, which in the future should be expanded to include more standard benchmark image sets, such as \textit{Set15}. Completing all of this future work will leave us well placed to expand our experiments to more models, including the current state of the art. If using FDPL to train these models results in even the modest performance increases seen in this paper, we will be able to achieve a new state of the art for the single image super resolution task.  

\bibliographystyle{IEEEbib}
\bibliography{refs}

\end{document}